\documentclass[prl,twocolumn,a4paper,superscriptaddress]{revtex4}
\usepackage{amsmath,epsfig,psfrag,natbib,bm}

\begin{document}


\newcommand{\Tr}{\text{tr}}
\newcommand{\lmax}{l_{\text{max}}}
\newcommand{\ra}{\rangle}
\newcommand{\la}{\langle}
\newcommand{\Bt}{\tilde{B}}


\newcommand{\mC}{\bm{C}}
\newcommand{\mU}{\bm{U}}
\newcommand{\mV}{\bm{V}}
\newcommand{\mW}{\bm{W}}
\newcommand{\mD}{\bm{D}}
\newcommand{\mI}{\bm{I}}
\newcommand{\mN}{\bm{N}}
\newcommand{\mS}{\bm{S}}


\newcommand{\mUt}{\tilde{\bm{U}}}
\newcommand{\mVt}{\tilde{\bm{V}}}
\newcommand{\mDt}{\tilde{\bm{D}}}


\newcommand{\vE}{\mathbf{E}}
\newcommand{\vB}{\mathbf{B}}
\newcommand{\vBt}{\tilde{\mathbf{B}}}
\newcommand{\vBw}{\mathbf{B}_W}
\newcommand{\vBwr}{\mathbf{B}_W^{(R)}}


\newcommand{\clp}{{\mathcal{P}}}


\title{Detecting magnetic CMB polarization on an incomplete sky}

\author{Antony Lewis}
 \email{Antony@AntonyLewis.com}
 \affiliation{DAMTP, CMS, Wilberforce Road, Cambridge CB3 0WA, UK.}
\author{Anthony Challinor}
 \email{A.D.Challinor@mrao.cam.ac.uk}
 \affiliation{Astrophysics Group, Cavendish Laboratory, Madingley Road, Cambridge CB3 OHE, UK.}
\author{Neil Turok}
 \email{N.G.Turok@damtp.cam.ac.uk}
 \affiliation{DAMTP, CMS, Wilberforce Road, Cambridge CB3 0WA, UK.}

\begin{abstract}
\vspace{\baselineskip}
The full sky cosmic microwave background polarization field can be
decomposed into `electric' and `magnetic' components.  Working in
harmonic space we construct magnetic variables that can be measured from
observations over only a portion of the sky. Our construction is exact for
azimuthally symmetric patches, but should continue to perform well for
arbitrary patches.
For isotropic, uncorrelated noise the variables have a very simple diagonal
noise correlation, and further analysis using them should be no harder than
analysing the temperature field.
We estimate the tensor mode amplitude that could be detected by the Planck
satellite and discuss the sensitivity of future experiments in the
presence of a contaminating weak lensing signal. 
\end{abstract}


\maketitle


Observations of fluctuations in the cosmic
microwave background (CMB) provide a wealth of information about
primordial inhomogeneities in the universe. 
One of the most interesting questions 
is whether there was a tensor (gravitational wave) component,
as predicted by simple inflationary models. CMB polarization 
measurements offer a unique probe of this
signal~\cite{Kamionkowski97,Zaldarriaga97,Hu98}. 

Polarization of the cosmic microwave sky is produced by electron
scattering, as photons decouple from the primordial plasma. 
Gravitational waves produce `magnetic' (i.e.\ curl) and `electric'
(i.e.\ gradient) polarization components at a comparable level
by anisotropic redshifting of the energy of photons. Magnetic
polarization is not produced by density perturbations, so
detection of a magnetic component would provide strong direct evidence
for the presence of a primordial gravitational wave (tensor) component.

The problem is
that the primordial polarization is only observable over the region
of the sky that is not contaminated by emission from our galaxy
and other foreground sources of polarization. 
But the electric/magnetic decomposition is inherently \emph{non-local},
and \emph{non-unique} in the presence of boundaries. To do an exact, lossless
separation on the incomplete sky one would need to know the boundary
conditions exactly---both the polarization \emph{and its derivative} around
the edge of the observed region. Clearly this is not feasible with noisy
observed data.

However the hypothesis of a magnetic signal 
can still be tested. One possibility which avoids
differentiating the observed data is to construct line integrals of the
polarization as discussed in Refs.~\cite{Chiueh01, Zaldarriaga01}.
The problem with these line integrals is that there
are an infinite number of them, and they are
not statistically independent. One would therefore prefer
a set of magnetic variables to which the 
electric component does not contribute,
but which are finite in number and
statistically independent (at least for idealized noise properties).
In this letter we shall construct such variables explicitly for
circular sky patches. A more complete exposition of our method using the
spin-weight formalism is given in Ref.~\cite{Lewis01}.

The observable polarization field is described in terms of the two
Stokes' parameters $Q$ and $U$ with respect to a particular choice of
axes about each direction on the sky. We use spherical polar
coordinates, and the Stokes' parameters define a
symmetric and trace-free (STF) rank two linear polarization tensor on
the sphere
\begin{equation}
\clp^{ab} = \frac{1}{2}[Q (\hat{\bm{\theta}} \otimes \hat{\bm{\theta}}
- \hat{\bm{\phi}} \otimes \hat{\bm{\phi}}) 
 - U (\hat{\bm{\theta}}\otimes\hat{\bm{\phi}}
+ \hat{\bm{\phi}}\otimes \hat{\bm{\theta}})]. 
\end{equation}
A two dimensional STF tensor can be written as a sum of
`gradient' and `curl' parts
\begin{equation}
\clp_{ab} = \nabla_{\la a}\nabla_{b\ra}P_E -
\epsilon^c{}_{(a}\nabla_{b)}\nabla_c P_B.
\end{equation}
where $\nabla$ is the covariant derivative on the sphere,
angle brackets denote the STF part on the enclosed indices, and
round brackets denote symmetrization.
The underlying scalar fields $P_E$ and $P_B$ describe
electric and magnetic polarization respectively and are clearly
non-local functions of the Stokes' parameters. One
can define scalar quantities which are local in the polarization by
taking two covariant derivatives to form $\nabla^a\nabla^b \clp_{ab}=
(\nabla^2+2)\nabla^2 P_E$ and
$\epsilon^{b}{}_c\nabla^c\nabla^a \clp_{ab} = (\nabla^2+2)\nabla^2 P_B$
which depend only on the electric and
magnetic polarization respectively. Here we focus on the magnetic
component, integrating the latter equation in order to construct an
observable that depends only on the magnetic component of $\clp_{ab}$,
but does not involve its derivatives. We start by
defining the surface integral
\begin{equation}
B_W' \equiv - 2\int_S \text{d}S \,W^* \epsilon^{b}{}_c\nabla^c\nabla^a
\clp_{ab},
\end{equation}
where $W$ is a window function defined over some patch $S$
of the observed portion of the sky. The factor of minus two is included to
make our definition equivalent to that in
Ref.~\cite{Lewis01}. Integrating by parts we have
\begin{eqnarray}
B'_W &=& \sqrt{2}\int_S \text{d}S\,   W_B^{ab}{}^* \clp_{ab} \nonumber\\
&-&
2\oint_{\partial S}
\text{d}l^a \left(\epsilon^b{}_a W^* \nabla^c \clp_{cb} -
\epsilon^b{}_c\nabla^c W^\ast \clp_{ab}\right),
\end{eqnarray}
where $W_{B\,ab}\equiv \sqrt{2}\epsilon^c{}_{(a} \nabla_{ b)} \nabla_{c} W$ is
an STF tensor window function.

We choose window functions so that the line
integrals that appear in the construction of $B'_W$  contain no
contribution from the  electric polarization. In general this requires that
$W$ and $\nabla_a W$ 
vanish on the boundary  $\partial S$. The surface integral
\begin{equation}
B_W \equiv \sqrt{2}\int_S \text{d}S \,  W^{ab}_B{}^* \clp_{ab}
\label{eq:ewbw}
\end{equation}
then depends only on the magnetic polarization.

We expand the window functions in spherical harmonics as 
\begin{equation}
W = \sum_{l \geq 2} \sum_{|m| \leq l} \sqrt{\frac{(l-2)!}{(l+2)!}}
W_{lm} Y_{lm}.
\label{eq:hwindow}
\end{equation}
(The square root factor is included for later convenience.) We need not
include $l=0$ and 1 spherical harmonics since they do not contribute to the
tensor window functions. In practice, we are only
interested in probing scales to some particular $\lmax$ (the
magnetic signal from tensor modes has maximal power for $l\approx 100$ and
decreases rapidly with $l$), so the sum in Eq.~(\ref{eq:hwindow}) can be
truncated at some finite $\lmax$.

The polarization tensor $\clp_{ab}$ can be expanded
over the whole sky in terms of STF tensor harmonics
\begin{equation}
\clp_{ab} = \frac{1}{\sqrt{2}} \sum_{lm} \left( E_{lm}\, Y_{(lm)ab}^G +
B_{lm}\, Y_{(lm)ab}^C\right),
\end{equation}
where $Y_{(lm)ab}^G$ and $ Y_{(lm)ab}^C$ are the gradient and curl tensor
harmonics of opposite parities defined in Ref.~\cite{Kamionkowski97}.
From the orthogonality of the spherical harmonics over the full sphere
it follows that
\begin{eqnarray}
B_{lm} = \sqrt{2}\int_{4\pi} \text{d}S\, Y_{(lm)}^{C\,ab*} \clp_{ab}.
\label{eq:EBlm}
\end{eqnarray}
In a rotationally-invariant ensemble, the expectation values of the harmonic
coefficients define the power spectrum:
\begin{equation}
\la B_{l'm'}^\ast B_{lm} \ra = \delta_{l'l}\delta_{m'm} C_l^{BB}.
\end{equation}

The form of the harmonic expansion~(\ref{eq:hwindow}) of the window function
ensures that the tensor window is
\begin{equation}
W_{B\, ab} = \sum_{lm} W_{lm} Y^C_{(lm)ab},
\end{equation}
where the sum is over $l\geq 2$ and $|m| \leq l$. Evaluating the surface
integrals in Eq.~(\ref{eq:ewbw}) we find
\begin{equation}
B_W = \sum_{lm} W_{lm}^\ast \tilde{B}_{lm}, 
\label{eq:ewbw_harm}
\end{equation}
where the pseudo-multipoles $\tilde{B}_{lm}$ are obtained by restricting the
integral in Eq.~\eqref{eq:EBlm} to the region $S$:
\begin{eqnarray}
\tilde{B}_{lm} = \sum_{l'm'} \left(W_{+(lm)(lm)'} B_{l'm'} - i W_{-(lm)(lm)'}
E_{l'm'}\right),
\end{eqnarray}
and the coupling matrices are given by
\begin{eqnarray}
W_{+(lm)(lm)'} &\equiv& \int_S \text{d}S\,Y^{C*}_{(lm)ab}
Y_{(lm)'}^{C\,ab} \\
W_{-(lm)(lm)'} &\equiv& i\int_S \text{d}S\,Y^{C*}_{(lm)ab}Y_{(lm)'}^{G\,ab}.
\end{eqnarray}
The matrix $W_{-(lm)(lm)'}$ controls the contamination 
with electric polarization and can always be written as
a line integral around the boundary of $S$. Our aim is to construct
window functions $W_{lm}$ that remove this contamination. For
azimuthally-symmetric patches the coupling matrices are block
diagonal ($W_{\pm (lm)(lm)'} \propto \delta_{mm'}$), and $W_{-(lm)(lm)'}$ has only two non-zero
eigenvalues for $|m|\ge 2$. For $m=0$ and $|m|=1$ there are zero and one
non-zero eigenvalues respectively.

We now give a practical method for constructing a non-redundant set
of window functions $\{ W_I \}$, where $I$ labels the particular window,
that achieve exact separation for azimuthal patches. The corresponding
(cleanly separated) magnetic observables will
be denoted $B_{W_I}$. Vectors are now denoted by bold Roman font,
e.g.\ $\vBw$ has components
$B_{W_I}$, and $\vB$ has components $B_{lm}$, and matrices are denoted by
bold italic font, e.g.\ $\mW_\pm$ have components $W_{\pm(lm)(lm)'}$.
We present the method in a form that is applicable (though no longer exact)
to arbitrary shaped regions $S$.

In matrix form we start with the observed data vector
\begin{equation}
 \vBt =  \mW_+ \vB  - i \mW_- \vE.
\end{equation}
The essence of our method is to choose a window
function $\mW$ so as to project out the range of $\mW_-$ corresponding
to its non-zero eigenvalues.

We first diagonalize $\mW_+$ so it can be written
$\mW_+ =\mU_+ \mD_+ \mU^\dagger_+$. This
allows us to identify the linear combinations $\mU^\dagger_+ \vB$ that are
poorly determined by $\vBt$---those
corresponding to the small diagonal elements of $\mD_+$. These
correspond to polarization patterns that are badly supported on the
observed patch of the sky and need to be removed from the analysis to
construct a non-redundant set of window functions.
The distribution of eigenvalues of $\mW_+$ is bimodal and the
exact definition of `small' is not 
critical~\cite{Lewis01}. 
We define an operator $\mUt_+$ which projects onto the eigenvectors of
$\mW_+$ whose eigenvalues are close to one. This amounts to removing
the appropriate columns of $\mU_+$ to give the column orthogonal matrix
$\mUt_+$. The corresponding smaller square diagonal matrix is $\mDt_+$,
and we have
\begin{equation}
\mUt^\dagger_+ \vBt \approx \mDt_+ \mUt^\dagger_+ \vB - i\mUt^\dagger_+
\mW_- \vE.
\end{equation}
We now multiply by $\mDt_+^{-1/2}$, defined by $[\mDt_+^{-1/2}]_{ij}
\equiv \delta_{ij}[ \mDt_+]_{ii}^{-1/2}$, to give
\begin{equation}
\mDt_+^{-1/2}\mUt^\dagger_+ \vBt \approx \mDt_+^{1/2} \mUt^\dagger_+
\vB - i\mDt_+^{-1/2}\mUt^\dagger_+ \mW_- \vE.
\end{equation}
This step ensures that isotropic noise in the map gives
isotropic noise on the variables $\mDt_+^{-1/2}\mUt^\dagger_+ \vBt$,
as shown in Ref.~\cite{Lewis01}.

We now project out the unwanted  $\vE$ contamination by projecting out of the
range of $\mDt_+^{-1/2}\mUt^\dagger_+ \mW_-$. The singular-value decomposition
for a non-square matrix takes the form $\mDt_+^{-1/2}\mUt^\dagger_+ \mW_- =
\mU\mD\mV^\dagger$ where $\mD$ is diagonal matrix, $\mU$ is unitary
and $\mV$ is a column orthogonal rectangular matrix. 
There are at most two non-zero singular values (diagonal elements of  $\mD$)
per $m$, and the corresponding left singular vectors (columns of $\mU$)
form an orthonormal basis for the range of $\mDt_+^{-1/2}\mUt^\dagger_+ \mW_-$.
We can project out of this range by defining $\mUt$ as the matrix obtained by
removing the columns of $\mU$ where the corresponding singular value
is non-zero. Thus, choosing the window functions
\begin{equation}
\mW^\ast = \mUt^\dagger \mDt_+^{-1/2}\mUt^\dagger_+,
\end{equation}
we obtain the vector of separated magnetic observables
\begin{eqnarray}
\vBw = \mW^\ast \vBt \equiv \mUt^\dagger \mDt_+^{-1/2}\mUt^\dagger_+ \vBt
\approx \mUt^\dagger \mDt_+ ^{1/2}\mUt^\dagger_+ \vB . \label{eq:bvar}
\end{eqnarray}
For azimuthal patches the separation is exact; the approximation sign
arises only from our use of $\mW_+ \approx \mUt_+ \mDt_+ \mUt_+^\dagger$
in simplifying the matrix that premultiplies $\vB$.

In the case of
idealized, un-correlated isotropic noise on the Stokes' parameters the
covariance of the noise $\Delta \vBw$ on $\vBw$ is simply~\cite{Lewis01} 
\begin{equation}
\mN \equiv \la \Delta \vBw \Delta \vBw^\dagger \ra = \sigma^2_N\mI,
\end{equation}
where $\sigma^2_N$ is the reciprocal weight per solid angle
on the Stokes' parameters. 

\begin{figure}[t!]
\begin{center}
\epsfig{figure=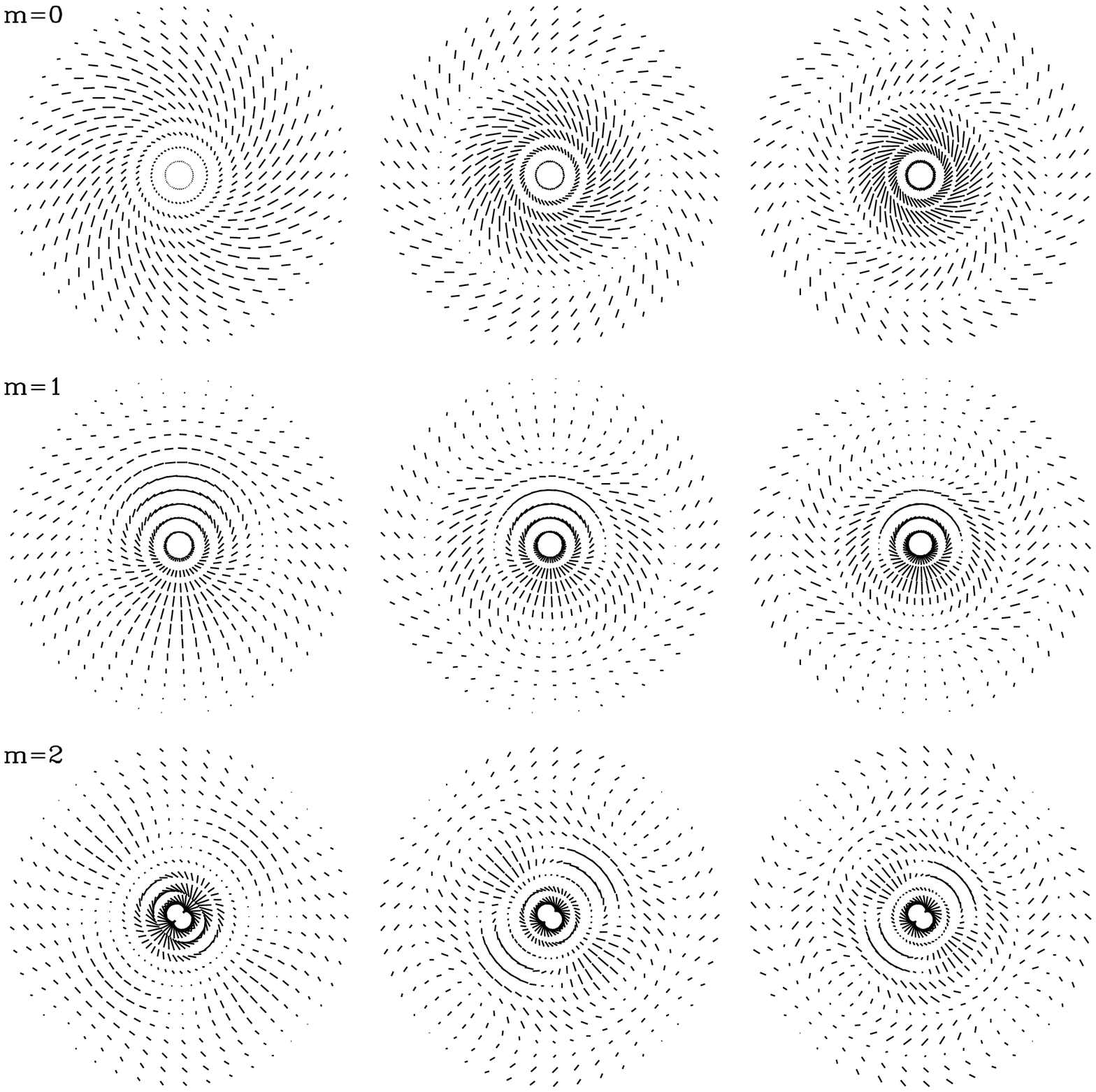,angle=0,width=8.5cm,clip=} 
\caption{The real space window functions for an azimuthally symmetric
sky patch with $\theta<10^\circ$, evaluated in the frame
where the signal is diagonal so the leftmost window produces the
largest signal for that $m$. 
\label{realspace}}
\end{center}
\end{figure}

We now consider how to test the null hypothesis that the
magnetic signal is due entirely to noise.
If the noise is Gaussian the $\vBw$ will be
Gaussian and the simplest thing to do is a $\chi^2$ test by computing
$\chi^2 = \vBw^\dag \mN^{-1} \vBw$ (for isotropic noise this is just
$\chi^2 = \vBw^\dag \vBw/\sigma^2_N$). Whilst the CMB magnetic polarization
signal is expected to be Gaussian, any spurious or unexpected signal
may not be. One may therefore also wish to do a more sophisticated set
of statistical tests at this point.

If the signal is Gaussian and we know the expected shape of the power
 spectrum we can use the statistic~\cite{Lewis01}
\begin{equation}
\nu'\equiv  \frac{\vBw^\dag \mN^{-1} \mS
\mN^{-1} \vBw - \Tr(\mN^{-1}\mS)}{\sqrt{4 \vBw^\dag \mN^{-1}\mS\mN^{-1}\mS\mN^{-1}\vBw
- 2\Tr(\mN^{-1}\mS\mN^{-1}\mS) }}.
\end{equation}
giving the number of standard deviations of the maximum likelihood
from zero for small signal to noise. Here the theory
covariance is
\begin{equation}
\mS \equiv  \la \vBw \vBw^\dag \ra = \mUt^\dagger\mDt^{1/2}_+\mUt^\dagger_+
\mC^{BB} \mUt_+ \mDt_+^{1/2} \mUt ,
\end{equation}
where the diagonal magnetic power spectrum matrix is given by
$[\mC^{BB}]_{(lm)(l'm')}=\delta_{m m'}\delta_{l l'}C_{l}^{BB}$. The
$\nu'$ distribution is computed 
by Monte-Carlo simulation in order to assign robust
probabilities~\cite{Lewis01}.

For a particular theoretical model we can rotate to the frame where the
signal matrix is diagonal, giving rotated $\vBwr$ that are fully
statistically independent. In Fig.~\ref{realspace} we plot the window
functions for the $\vBwr$ which give the largest contributions to the
signal for a typical flat scale-invariant $\Lambda$CDM model.
The window functions
are plotted as line segments of length $\sqrt{Q_W^2 + U_W^2}$ at angle
$\tan^{-1}(U_W/Q_W)/2$ to the $\theta$ direction where the real 
quantities $Q_W$ and $U_W$ are defined so that
\begin{equation}
\Re B_W^{(R)} = \int_S \text{d}S\left( Q_W Q + U_W U\right).
\end{equation}
%

\begin{figure}[t!]
\begin{center}
\epsfig{figure=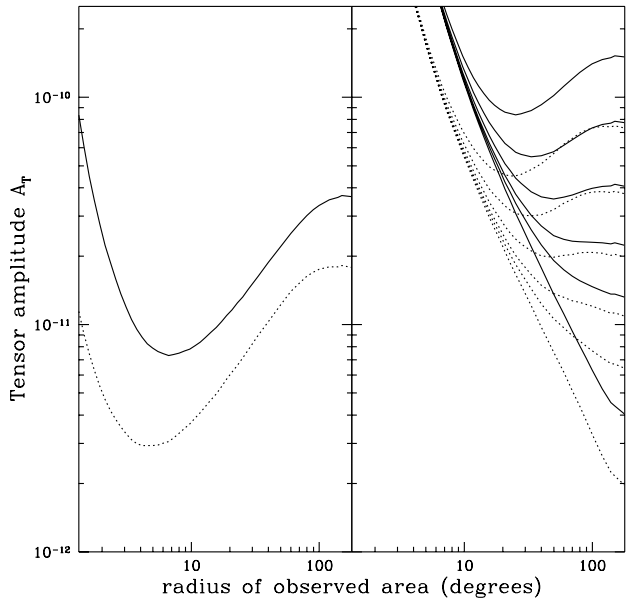,angle=0,width=8.5cm} 
\caption{The smallest gravitational wave amplitude $A_T$ (defined as
in Ref.~\cite{Martin00;long}) that could be detected at $99$ per cent
confidence with probability $0.95$ (solid lines) and $0.5$ (dotted
lines) by a one year survey that maps a circular patch of sky of a given
radius assuming uniform noise. The left panel is for detector
sensitivity $s= 10\mu \text{K}\sqrt{\text{sec}}$ (about three times
better than Planck) and no lensing signal, the right panel is for sensitivities (bottom to top) $s^2= \{0, 25, 50,
100, 200, 400\}\,\mu\text{K}^2\text{sec}$ and a
magnetic lensing signal with white spectrum $C^{BB}_{\text{lens}} = 4.4\times 10^{-6}\,\mu\text{K}^2$.
\label{patchsize}}
\end{center}
\end{figure}

Of the current funded experiments, only Planck is likely to detect magnetic
polarization if the levels are as predicted by standard cosmological models.
As a toy model we consider the $143$ and $217$ GHz polarized channels of the
Planck High Frequency Instrument. We approximate the noise as isotropic
and ignore the finite beam width. Combining maps from these
two channels with inverse variance weighting, we find $\sigma_N \approx
6 \times 10^{-3} \mu \text{K}/\text{K}$, where $Q$ and $U$ are expressed
as dimensionless thermodynamic equivalent temperatures in units of the
CMB temperature. We apply an azimuthally-symmetric galactic cut of
$20$ degrees either side of the equator and use $\lmax=250$. We find
that the null-buster would give a detection at $99$ per cent significance for a
tensor initial power spectrum amplitude (defined as in
Ref.~\cite{Martin00;long}) $A_T \agt  4\times 10^{-10}$ with
probability greater than $90$ per cent. This corresponds to about 1/10
of the large scale $C_l$ detected by COBE, and would be generated by inflationary
models with slow-roll parameter $\epsilon \agt 0.01$ and energy scale
$V^{1/4}\agt 2\times 10^{16}\,\text{GeV}$ at horizon crossing (for
example the simplest $\phi^2$ inflation model). 

Considering experiments more sensitive than Planck, one must also take
into account the magnetic signal produced by
weak lensing of the much larger electric polarization~\cite{Zaldarriaga98,Guzik00}. 
We model this lensing signal as an additional constant Gaussian white
noise.
In Fig.~\ref{patchsize} we show the minimum value of $A_T$ that could
be detected at $99$ per cent significance
as a function of the radius of the
observed patch, assuming a flat $\Lambda$CDM model which reionizes at
$z=6.5$ (as evidenced by recent Keck observations).  More sensitive surveys require larger patches in order to
distinguish the tensor signal from the lensing signal,
and there is an optimal survey size when the instrument noise is
approximately equal to the lensing noise. Even with no instrument
noise the lensing
signal places a lower limit on what can be detected, and in this limit
one should survey as much of the sky as possible.
Since we have assumed that source subtraction and component
separation can be done exactly our results are probably optimistic.

In summary, we have shown how to construct a set of statistically
independent magnetic polarization observables from observations covering
only a portion of the sky.  For large patches these contain most of the
available information (see
Ref.~\cite{Lewis01} for a quantitative discussion). Some information
loss is inevitable as one cannot do lossless clean separation without
taking derivatives of the polarization on the boundary. The
variables have simple noise properties and can be
constructed easily using our harmonic-based approach that automatically
removes redundancy due to the finite sky coverage. 

\begin{acknowledgments}
AL and NT are supported by PPARC. AC acknowledges a PPARC Postdoctoral
Fellowship.
\end{acknowledgments}

\vspace{\baselineskip}



\end{document}